\documentclass[11pt]{article}
\usepackage[utf8]{inputenc}

\usepackage{cite}

\usepackage{nameref,hyperref}
\usepackage{eucal}
\usepackage{amsmath,amssymb}

\usepackage[right]{lineno}

\usepackage{microtype}
\usepackage{changepage}

\usepackage[aboveskip=1pt,labelfont=bf,labelsep=period,singlelinecheck=off]{caption}

\usepackage{placeins}

\usepackage{graphicx}
\usepackage{amsmath}
\usepackage[version=4]{mhchem}
\usepackage{siunitx}
\usepackage{longtable,tabularx}
\usepackage{authblk}

\usepackage{subcaption} % for subfigures

\title{Smart Acoustic Lining for UHBR Technologies Engine Part 1: design of an electroacoustic liner and experimental
	characterization under flow in rectangular cross-section ducts}

\author{E. De\;Bono\footnote{Post-doctoral researcher, LTDS \'Ecole Centrale de Lyon, emanueledeb88hotmail.it.}}

\author{M. Collet\footnote{Research Director, LTDS \'Ecole Centrale de Lyon, manuel.collet@ec-lyon.fr.}}
\author{K. Billon\footnote{Research Engineer, LTDS \'Ecole Centrale de Lyon, kevin.billon@ec-lyon.fr.}}
\affil{Univ Lyon, CNRS, \'Ecole Centrale de Lyon, LTDS, UMR5513, 69130 Ecully, France.}

\author{E. Salze\footnote{Research Engineer, LMFA \'Ecole Centrale de Lyon, edouard.salze@ec-lyon.fr.}}
\affil{Univ. Lyon, \'Ecole Centrale de Lyon, LMFA, UMR 5509, F-69134 Ecully, France.}

\author{H. Lissek\footnote{Research Director, LTS2 \'Ecole Polytechnique Fédérale de Lausanne, herve.lissek@epfl.ch.}}
\author{M. Volery\footnote{Post-doctoral researcher, LTS2 \'Ecole Polytechnique Fédérale de Lausanne, maxime.volery@epfl.ch.}}
\affil{Signal Processing Laboratory LTS2, \'Ecole Polytechnique Fédérale de Lausanne, Station 11,	CH-1015 Lausanne, Switzerland.}

\author{M. Ouisse\footnote{Professor, DMA, Université de Franche-Comté, morvan.ouisse@femto-st.fr.}}
\affil{SUPMICROTECH, Université de Franche-Comté, CNRS, institut FEMTO-ST, F-25000 Besançon, France.}

\author{J. Mardjono\footnote{Research Director, Safran Aircraft Engines, jacky.mardjono@safrangroup.com.}}
\affil{Safran Aircraft Engines, F-75015, Paris, France.}

\begin{document}

\maketitle

\begin{abstract}
The new generation of Ultra-High-By-Pass-Ratio (UHBR) turbofan engine while considerably reducing fuel consumption, threatens higher noise levels at low frequencies because of its larger diameter, lower number of blades and rotational speed. This is accompanied by a shorter nacelle, leaving less available space for acoustic treatments. In this context, a progress in the liner technology is highly demanded, prospecting alternative solutions to classic liners. The SALUTE H2020 project has taken up this challenge, proposing electro-active acoustic liners, made up of loudspeakers (actuators) and microphones (sensors). The electro-active means allow to program the surface impedance on the electroacoustic liner, but also to conceive alternative boundary laws. Test-rigs of gradually increasing complexities have allowed to raise the Technology Readiness Level (TRL) up to 3-4. In this first part, we describe the control laws employed in the experimental campaigns, and present the scattering performances in rectangular waveguides with monomodal guided propagation. These results have assessed the isolation capabilities, stability and robustness of such programmable boundary technology, allowing to gain confidence for the successive implementation in a scaled turbofan test-rig.
\end{abstract}

%\section{Nomenclature}

%{\renewcommand\arraystretch{1.0}
%	\noindent\begin{longtable*}{@{}l @{\quad=\quad} l@{}}
%		$t$  & time variable \\
%		$\partial t$ &   partial derivative with respect to time\\
%		$p$& sound pressure \\
%		$v_n$ & normal velocity on the acoustic liner, directed inward the liner\\
%		$Z_{Loc}(\partial t)$ & local impedance differential operator in time domain \\
%		$U_b$ & advection boundary speed \\
%		$\partial_x$ & partial derivative along the longitudinal direction in the downstream sense\\
%		$IL$ & Insertion Loss for grazing incidence\\
%		$R$ & Reflection coefficient for grazing incidence\\
%		$\alpha$ & absorption coefficient for grazing incidence\\
%\end{longtable*}}

\section{Introduction}
The acoustic problem of interest here, is the noise transmission mitigation in an open duct, by treatment of the parietal walls with the so-called liners. Examples of industrial fields where this problem is particularly felt are the Heating and Ventilation Air-Conditioning Systems (HVAC) and the aircraft turbofan engines. The new generation of Ultra-High-By-Pass-Ratio (UHBR) turbofans, in order to comply with
the significant restrictions on fuel consumptions and pollutant emissions, present larger diameter, lower number of blades and rotational speed and a shorter nacelle. These characteristics conflict with the equally restrictive regulations on noise pollution, as the noise signature is shifted toward lower frequencies, which are much more challenging to be mitigated by parietal treatments. The acoustic liner technology applied nowadays for noise transmission attenuation at the inlet and outlet portions of turbofan engines is the so-called Single or Multi-Degree-of-Freedom liner, whose working principle relates to the quarter-wavelength resonance, and demands larger thicknesses to target lower frequencies. They are made of a closed honeycomb structure and a perforated plate which is used to provide the dissipative effect, to add mass in order to decrease the resonance frequency, and also to maintain the aerodynamic flow as smooth as possible on the internal wall of the nacelle. As the honeycomb structure is impervious, propagation is prevented transversely to the wall, therefore it can be considered as \emph{locally reacting} as long as the incident field wavelength is much larger than the size of the honeycomb cells \cite{ma2020development}.\\
A first interest for active control is the possibility to tune the resonators to different frequencies. Many adaptive Helmholtz resonator solutions have been proposed by varying either the acoustic stiffness (i.e. the cavity as in \cite{hermiller2013morphing}), or the acoustic mass (i.e. the orifice area, as in \cite{esteve2004development}), but both these techniques tended to present complex structure, excessive weight and high energy consumption \cite{ma2020development}.\\
Active Noise Cancellation (ANC) have provided alternative solutions for achieving higher attenuation levels. From the seminal idea of Olson and May \cite{olson1953electronic}, first active \emph{impedance control} strategies \cite{guicking1984active,galland2005hybrid} proposed an ``active equivalent of the quarter wavelength resonance absorber'' for normal and grazing incidence problems, respectively. The same technique was slightly modified by \cite{betgen2011new} in the attempt to reproduce the Cremer's liner optimal impedance for the first duct modes pair \cite{cremer1953theory,tester1973optimization}. As such impedance could not be achieved in a broadband sense, this approach remained limited to monotonal applications.\\
These are examples of impedance control achieved through secondary source approaches combined with passive liners, but the collocation of sensor and actuator suggested also another avenue: the modification of the actuator (loudspeaker or else) own mechano-acoustical impedance. The objective shifts from creating a ``quite zone'' at a certain location, to achieving an optimal impedance on the loudspeaker diaphragm.
This is the Electroacoustic Resonator (ER) idea, which have found various declinations, such as electrical-shunting \cite{fleming2007control}, direct-impedance control \cite{furstoss1997surface} and self-sensing \cite{leo2000self}. In order to overcome the low-flexibility drawback of electrical shunting techniques, minimize the number of sensors, meanwhile avoiding to get involved into the electrical-inductance modelling of the loudspeaker, a pressure-based current-driven architecture proved to achieve the best absorption performances in terms of both bandwidth and tunability \cite{rivet2016broadband}. It employs one or more pressure sensors (microphones) nearby the speaker, and a model-inversion digital algorithm to target the desired impedance by controlling the electrical current in the speaker coil. Compared to classical ANC strategies, the impedance control is conceived to assure the acoustical passivity of the treated boundary, and hence the stability of the control system independently of the external acoustic environment \cite{goodwin2001control}. The correlation between passivity and stability of the ER, has been analysed in \cite{de2019electroacoustic,de2022effect}. The same architecture of ER has also been exploited to conceive an alternative control algorithm capable of enforcing nonlinear acoustical responses of the ER at low excitation levels \cite{DeBono2024,MORELL2024118437,da2023experimental,morell2023control,DeBono2022spie,morellFA2023nonlinear}.
In \cite{collet2009active}, for the first time, a boundary operator involving the spatial derivative was targeted by distributed electroacoustic devices. It was the first form of the Advection Boundary Law (ABL), then implemented on ER arrays lining an acoustic waveguide in \cite{KarkarDeBono2019}, where it demonstrated non-reciprocal sound propagation. The performances of such \emph{generalized impedance} control law have been studied both in its local and nonlocal declination in \cite{de2024advection,de2023nonlocal,salze2023electro,billon2022flow,billon2021experimental,billon20222d,billon2023smart}.\\
The SALUTE projects has the objective to prompt the Technology-Readiness-Level (TRL) of a liner made up of ERs. Compact cells, each one integrating 4 microphones, a speaker and a microprocessor to execute the control algorithm, compose an array which can replace the parietal walls of a duct. In order to protect the electromechanical devices from the turbolent flow, a wiremesh, sustained by a perforated plate, is placed in front of the electroacoustic liner. Such protective liner is supposed to be quasi-transparent with respect to the acoustic field. In this first part of our contribution, we analyse the results obtained by the numerical and experimental campaign to characterize the performances of the electroacoustic liner in presence of flow. The importance of these achievements rely in the assessment of such innovative boundary control technology, for noise isolation in acoustic waveguides with flow. We report here the results achieved at two intermediate test-bench, which allowed to gain the necessary confidence for testing our liner in a scaled 1:3 reproduction of turbofan engine, called PHARE-2 \cite{pereira2019new}. The experimental campaign on PHARE-2 follows in the Part 2 of this work.

\section{Numerical simulations without flow}

First of all, let us define the boundary operator we target by our electroacoustic liner. It is called Advection Boundary Law (ABL), which writes:

%\begin{equation}\label{eq:advection law in vy with Z}
%	\begin{split}
%		Z_{Loc}(\partial_t) \ast \partial_tv = \partial_t p + U_b\partial_x p \;\;\;\;\textrm{on $\partial\Omega$}.\\
%	\end{split}
%\end{equation}

\begin{equation}\label{eq:advection law in vy with Z}
	\begin{split}
		Z_{Loc}(\partial_t) \ast \partial_tv_n = \partial_t p + U_b\partial_x p,\\
	\end{split}
\end{equation}

where $Z_{Loc}(\partial_t)$ is a Single-Degree-Of-Freedom (SDOF) local impedance convolution operator in time, $\ast$ is the convolution operation, $v_n$ is the normal velocity at the boundary, $p$ the pressure. For classical locally reacting liners, $Z_{Loc}(\partial_t)$ relates the local acceleration to the time derivative of the local sound pressure $\partial_tp$. Our \emph{generalized impedance} operator includes an additional term, given by $U_b$ times the spatial gradient of pressure. $U_b$ is the artificial advection speed introduced at the boundary. We can define $M_b=U_b/c_0$. Notice that for $M_b=0$, we retrieve a classical SDOF local impedance.

\begin{figure}[ht!]
	\centering
	\includegraphics[width=\textwidth]{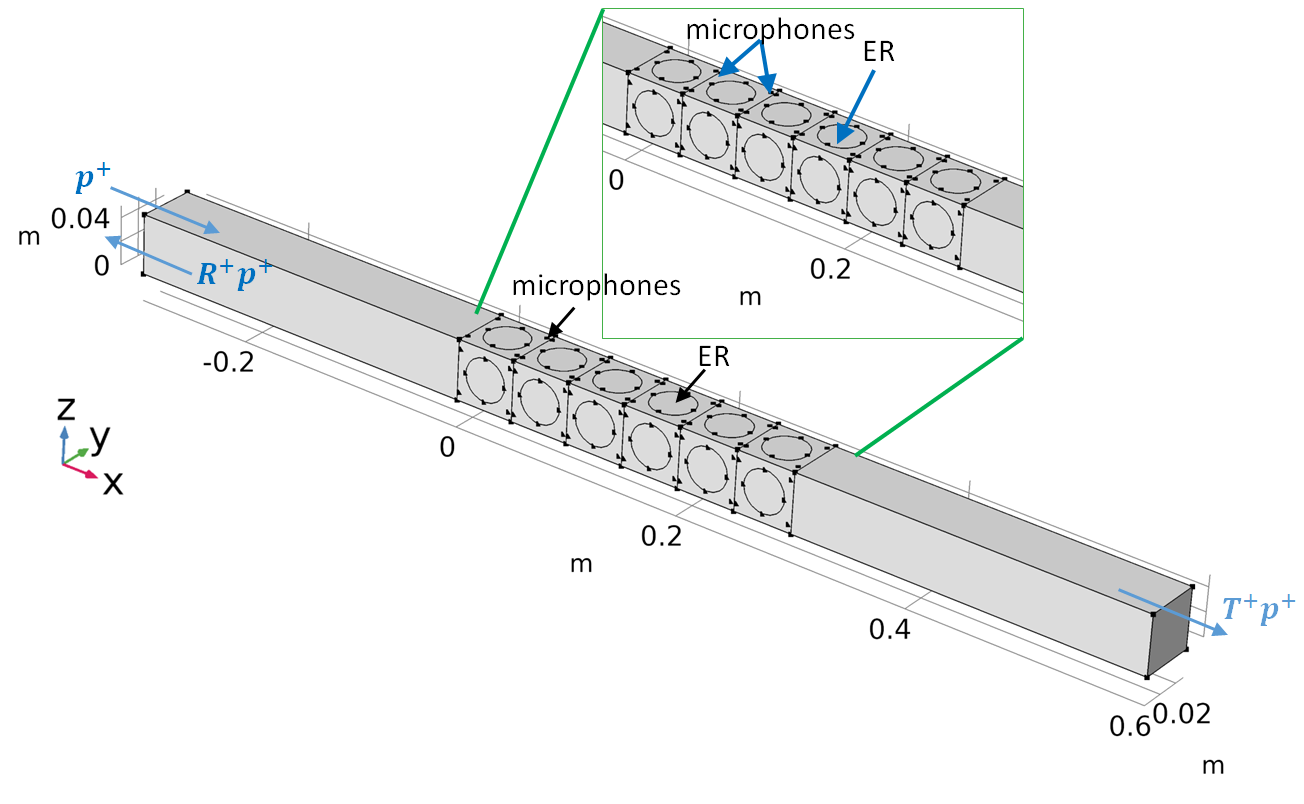}
	\centering
	\caption{3D geometry for scattering simulations, in case of ERs disks applied flush on the duct boundary.}
	\label{fig:duct3D_punctulamicros_LSinfront_0.05d_sect}
\end{figure}

\begin{figure}
	\centering
	\includegraphics[width=.6\textwidth]{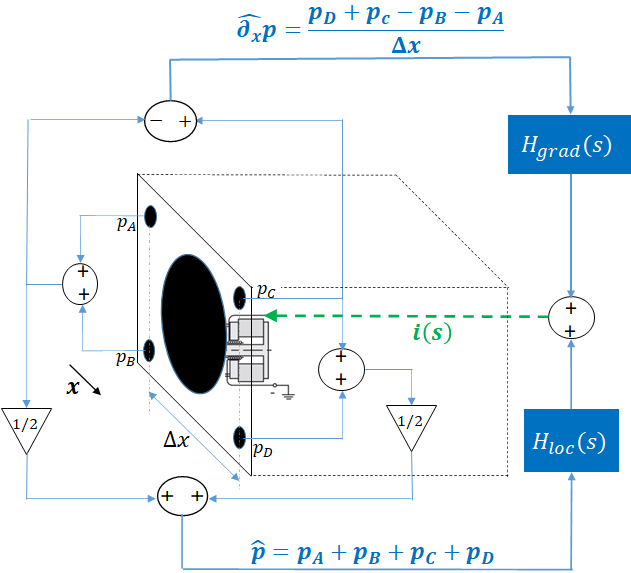}\\
	\caption{Sketch of the 4-microphones ER control, corresponding to Eq. \eqref{eq:controlling current}.}
	\label{fig:loudspeaker_with_NonLocalControl_scheme}
\end{figure}

In this section we simulate the scattering performances in the plane wave regime of a 3D acoustic waveguide, of square cross section with $5$ cm lateral sides, without flow. The ABL is applied along each side of the duct for a length of $30$ cm. In order to investigate the effect of discretizing the ABL by individual ERs lining the parietal walls of a rectangular cross section duct, as in the experimental test-rig of Section \ref{sec:first testbench}, the ABL is applied on separate disks of diameter $3.6$ cm, simulating the ERs (6 per each duct edge), as showed in Fig. \ref{fig:duct3D_punctulamicros_LSinfront_0.05d_sect}. The dynamics of each speaker is simulated according to the Thiele-Small SDOF model \cite{beranek2012acoustics}.

%The time delay, indeed, significantly impacts the acoustical passivity of the ER at high frequencies \cite{de2022effect}. Therefore, it is interesting to visualize its effects also on the grazing-incidence scattering coefficients. \\%%%%%%%%%%%%%%%%%%%%%%%%%%%%%%%%%%%%%%%%%%%%%%%%%%%%%

The loudspeaker model is reported in Eq. \eqref{eq:loudspeaker model}, in terms of the Laplace variable $s$:

\begin{equation}\label{eq:loudspeaker model}
	Z_0(s) \bar{v}(s) = \bar{p}(s) - \frac{Bl}{S_e}\bar{i}(s).
\end{equation}

In Eq. \eqref{eq:loudspeaker model}, $\bar{p}(s)$ and $\bar{v}(s)$ are the acoustic pressure and velocity, respectively, on the speaker diaphragm,  $\bar{i}(s)$ is the electrical current in the speaker coil, $Z_0(s)=M_0s+R_0+K_0/s$ is the acoustical impedance of the loudspeaker in open circuit, with $M_0$, $R_0$ and $K_0$ the corresponding acoustical mass, resistance and stiffness. The electrical current $\bar{i}(s)$ is multiplied by the force factor $Bl$ to get the electromagnetic force, and divided by the effective area $S_e$ to retrieve an equivalent pressure.
The ABL is implemented by defining the electrical current $i(s)$ as in Eq. \eqref{eq:controlling current}:

\begin{equation}
	\label{eq:controlling current}
	i(s)  = H_{loc}(s)\hat{\bar{p}}(s) + H_{grad}(s)\hat{\partial_x} \bar{p}(s),
\end{equation}

where $\hat{\bar{p}}(s)$ and $\hat{\partial_x} \bar{p}(s)$ are the estimated local pressure and its x-derivative on each speaker diaphragm, in the Laplace domain. The local sound pressure is estimated by averaging the four microphones on the corners of each ER $\hat{p}=(p_A+p_B+p_C+p_D)/4$, while the x-derivative is estimated by a first-order finite difference $\hat{\partial_x} p=\biggr((p_C+p_D)-(p_A+p_B)\biggr)/\Delta x$, with $\Delta x\approx 4$ cm the distance between the microphones before (A,B) and after (C,D) each ER speaker, along the x-direction, as showed in Fig. \ref{fig:loudspeaker_with_NonLocalControl_scheme}. A time delay of $\tau=2\times10^{-5}$ seconds between the pressure inputs and the electrical current, is considered by multiplying the microphones pressures by $e^{-\mathrm{j}\omega\tau}$, in order to simulate the physiological latency of the digital control algorithm of the ER \cite{de2022effect}.\\
The transfer functions $H_{loc}(s)$ and $H_{grad}(s)$ are defined based upon the loudspeaker model of Eq. \eqref{eq:loudspeaker model}. Equating the velocity of the speaker diaphragm from Eq. \eqref{eq:loudspeaker model}, and the velocity corresponding to the ABL (Eq. \eqref{eq:advection law in vy with Z}), we get the expressions in the Laplace space of $H_{loc}$ and $H_{grad}$, in Eq.s \eqref{eq:Hloc} and \eqref{eq:Hgrad}, respectively.

\begin{equation}
	\label{eq:Hloc}
	H_{loc}(s) = \frac{S_e}{Bl} \biggr(1 - \frac{Z_{0}(s)}{Z_{Loc}(s)}\biggr),\\
\end{equation}

\begin{equation}
	\label{eq:Hgrad}
	H_{grad}(s) = -\frac{S_e}{Bl}\frac{Z_{0}(s)}{Z_{Loc}(s)}\frac{U_b}{s}F_{hp}(s),\\
\end{equation}

where $F_{hp}(s)$ in $H_{grad}(s)$ is a high-pass filter necessary in order for $H_{grad}(\mathrm{j}\omega)$ not to become infinite for $\omega\to 0$.
The synthesis of our corrector transfer functions, is also called \emph{model inversion} \cite{devasia2002should} approach, as the objective of the controller is to cancel out the loudspeaker proper dynamics, and replace it with a desired acoustic behaviour.\\
Both Eq.s \eqref{eq:Hloc} and \eqref{eq:Hgrad} are implemented in the Comsol model. From the microphones estimation of $\hat{p}$ and $\hat{\partial_x} p$, the electrical current $i$ is obtained from Eq. \eqref{eq:controlling current}. Hence, the loudspeaker dynamics Eq. \eqref{eq:loudspeaker model} is solved for $\bar{v}(s)$, which is then imposed on the disks representing the speaker membranes in the numerical model.\\

\begin{figure}
	\centering
	\includegraphics[width=.6\textwidth]{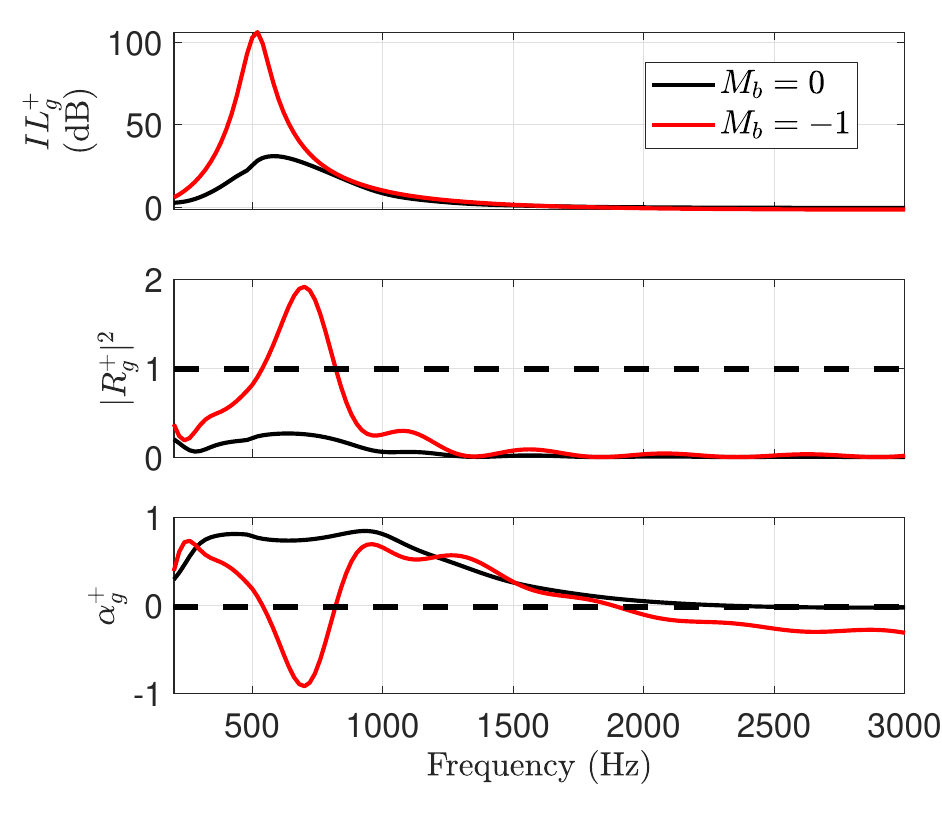}\\
	\caption{Comparison between local impedance control ($M_b=0$) and ABL ($M_b=-1$), in terms of scattering coefficients in the 3D waveguide.}
		\label{fig:Scatt3D_NominalCase_0.5muKmuM_1rho0c0Rat_-1cac0_2e-05tau_OCvsLocalvsNonLocal}
\end{figure}

In Fig. \ref{fig:Scatt3D_NominalCase_0.5muKmuM_1rho0c0Rat_-1cac0_2e-05tau_OCvsLocalvsNonLocal}, the scattering coefficients achieved by the ABL with $M_b=-1$, are plotted along with the ones relative to local impedance control ($M_b=0$), applied on each ER. The ABL demonstrates higher isolation capabilities, though being non-passive slightly after resonance. Notice also the loss of passivity at high frequencies (above 2 kHz). This is mostly due to the time delay \cite{de2022effect}.

\FloatBarrier

\section{First test bench without flow}\label{sec:first testbench}

\begin{figure}[ht!]
	\centering
	\includegraphics[width=0.7\textwidth]{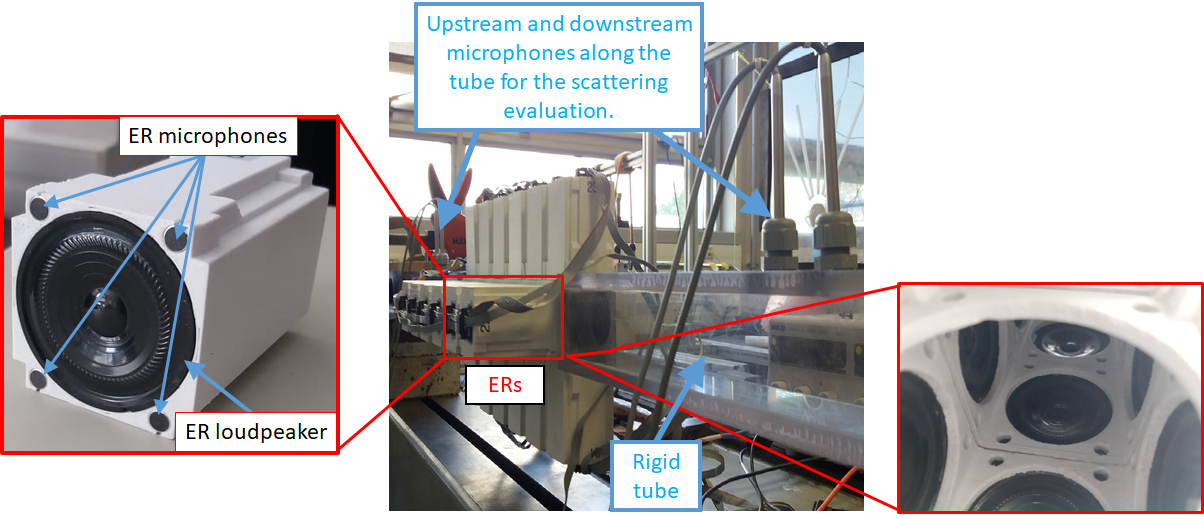}
	\caption{ER prototype (left); waveguide (middle) for the scattering evaluation, with internal view of the lined segment (right).}
	\label{fig:TL_ExpSetup}
\end{figure}

%\begin{figure}[ht!]
%	\centering
%	\includegraphics[width=0.7\textwidth]{TL_Setup_Sketch.png}
%	\caption{Sketch of the test-bench.}
%	\label{fig:TL_Setup_Sketch}
%\end{figure}

\begin{figure}[ht!]
	\centering
	\includegraphics[width=0.5\textwidth]{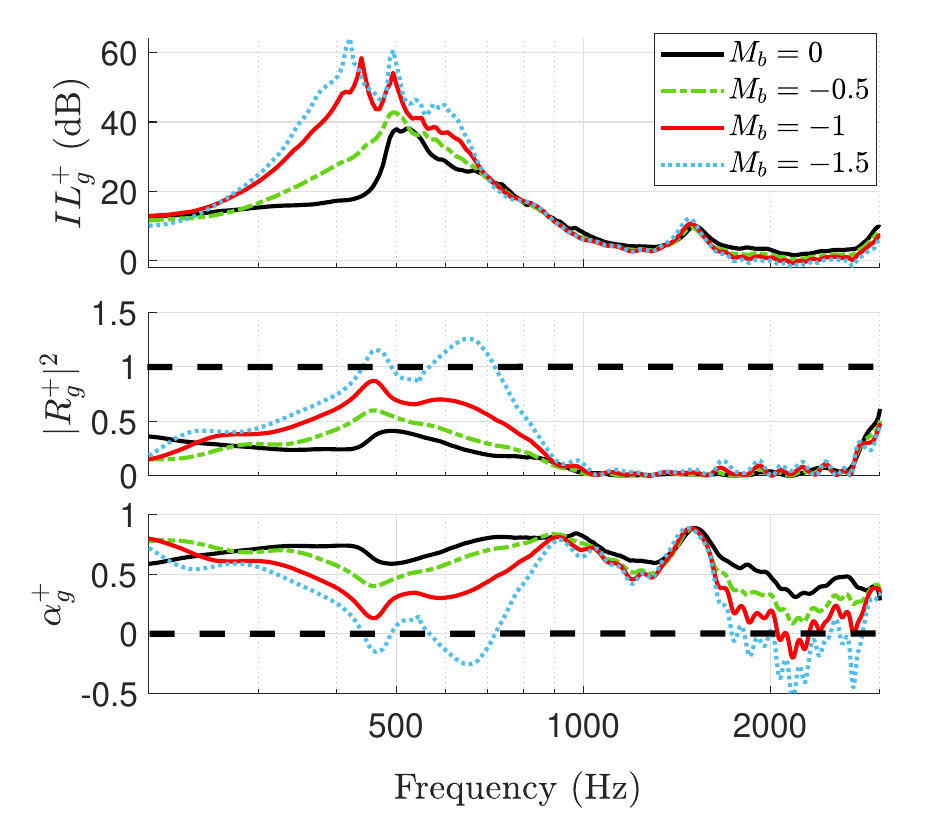}
	\centering
	\caption{Experimental scattering performances for incident field propagating toward $+x$, achieved by the ABL with varying $M_b$.}
	\label{fig:ExpSCATTplus_VARYINGcac0_0.5muM_1rho0c0Rat_0.5muK}
\end{figure}

In this section, the advection control law is experimentally tested on an array of 24 ER prototypes lining a squared cross-section duct of about $0.05$ m side, as illustrated in the photos of Fig. \ref{fig:TL_ExpSetup}. The ERs are placed 6 per each side of the duct, as showed in Fig. \ref{fig:TL_ExpSetup}. Each ER has a surface area of about $0.05\times0.05$ $m^2$, for a total lined segment length of about $0.3$ m in the duct.\\
The four scattering coefficients have been estimated according to the two-source method \cite{munjal1990theory}. Fig. \ref{fig:ExpSCATTplus_VARYINGcac0_0.5muM_1rho0c0Rat_0.5muK} shows the experimental scattering coefficients for incident field toward $+x$, with varying $M_b$. Fig. \ref{fig:ExpSCATTplus_VARYINGcac0_0.5muM_1rho0c0Rat_0.5muK} confirms the higher isolation achieved by increasing the absolute value of $M_b<0$. Observe, in Fig. \ref{fig:ExpSCATTplus_VARYINGcac0_0.5muM_1rho0c0Rat_0.5muK}, the reduction of passivity from 1.8 kHz and above with higher $|M_b|$. This is due to a combined effect of time delay and the first order approximation of $\hat{\partial}_x p$, which is clearly amplified for higher values of $|M_b|$.

%Each ER is controlled autonomously, and the control architecture is illustrated in Fig. \ref{fig:EA_with_Howland}: the signals $\hat{p}$ and $\hat{\partial}_xp$ on the speaker diaphragm, after being digitally converted by the Analogue-Digital-Converter (ADC), are fed into a \emph{programmable} digital signal processor (DSP) where the output of the control is computed at each time step. The Howland current pump \cite{pease2008comprehensive} allows to enforce the electrical current $i$ in the speaker coil independently of the voltage at the loudspeaker terminals. It consists of an operational amplifier, two input resistors $R_i$, two feedback resistors $R_f$, and a current sense resistor $R_s$. The resistance $R_d$ and capacitance $C_f$ constitutes the compensation circuit to ensure stability with the grounded load \cite{steele1992tame}. More details can be found in \cite{de2022effect}.
%
%\begin{figure}
%	\centering
%	\includegraphics[width=.6\textwidth]{EA_with_Howland.png}
%	\caption{Sketch of the ER architecture.}
%	\label{fig:EA_with_Howland}
%\end{figure}
%All ERs and control interfaces have been produced in the Department of Applied Mechanics at FEMTO-st Institute. The control laws have already been defined in Section \ref{sec:scatt 3D}, by Eq. \eqref{eq:controlling current}, \eqref{eq:Hloc}, \eqref{eq:Hgrad}, and the loudspeaker parameters provided in Table \ref{tab:TSparam FEMTO}. 

\section{Caiman test-bench: monomodal propagation with flow}

\begin{figure}[ht!]
	\centering
	\includegraphics[width=0.95\textwidth]{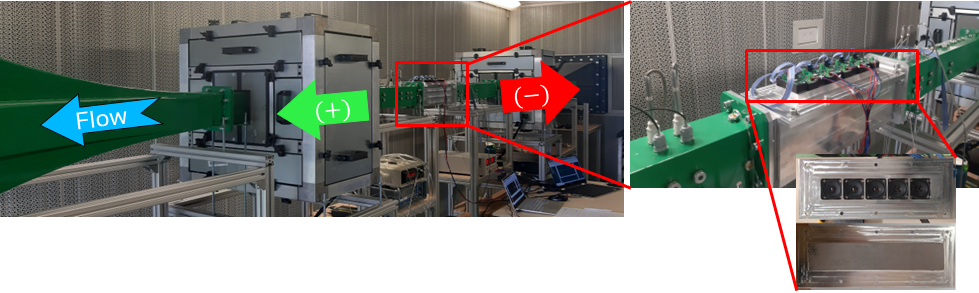}
	\caption{Left: ``Caiman'' wind tunnel, available in the Laboratory of Fluid Mechanics and Acoustics in the Ecole Centrale de Lyon. The downstream and upstream sense of propagation are indicated by the green and red arrows respectively. Right: zoom on the treated section with our electroacoustic liner covered by a wiremesh.}
	\label{fig:Caiman_Testrig}
\end{figure}

Here we report the results of the experimental campaign on the test-bench Caiman, in the Laboratory of Fluid Mechanics and Acoustics (LMFA) of the \'Ecole Centrale de Lyon. In the operative frequency range of our electroacoustic liner, the Caiman duct exhibits monomodal propagation (plane waves in the rigid segments). A connected wind-tunnel allows to reach up to Mach 0.3 in the waveguide. Appropriate noise sources are placed on both sides of the lined segment, in order to solve the full scattering problem according to the two-source method \cite{munjal1990theory}. The experimental test-bench is reported in Fig. \ref{fig:Caiman_Testrig}. On the left of Fig. \ref{fig:Caiman_Testrig}, the entire waveguide is visible, with the two grey boxes containing the noise sources on each side of the lined segment, and the indication of the downstream ($+$) and upstream ($-$) senses of propagation. On the right of Fig. \ref{fig:Caiman_Testrig}, the lined segment is showed, comprising 6 ERs covered by a wiremesh sustained by a perforated plate to protect the electromechanical devices from the flow. 

\begin{figure}[ht!]
	\centering
	\begin{subfigure}[ht!]{0.45\textwidth}
		\centering
		\includegraphics[width=\textwidth]{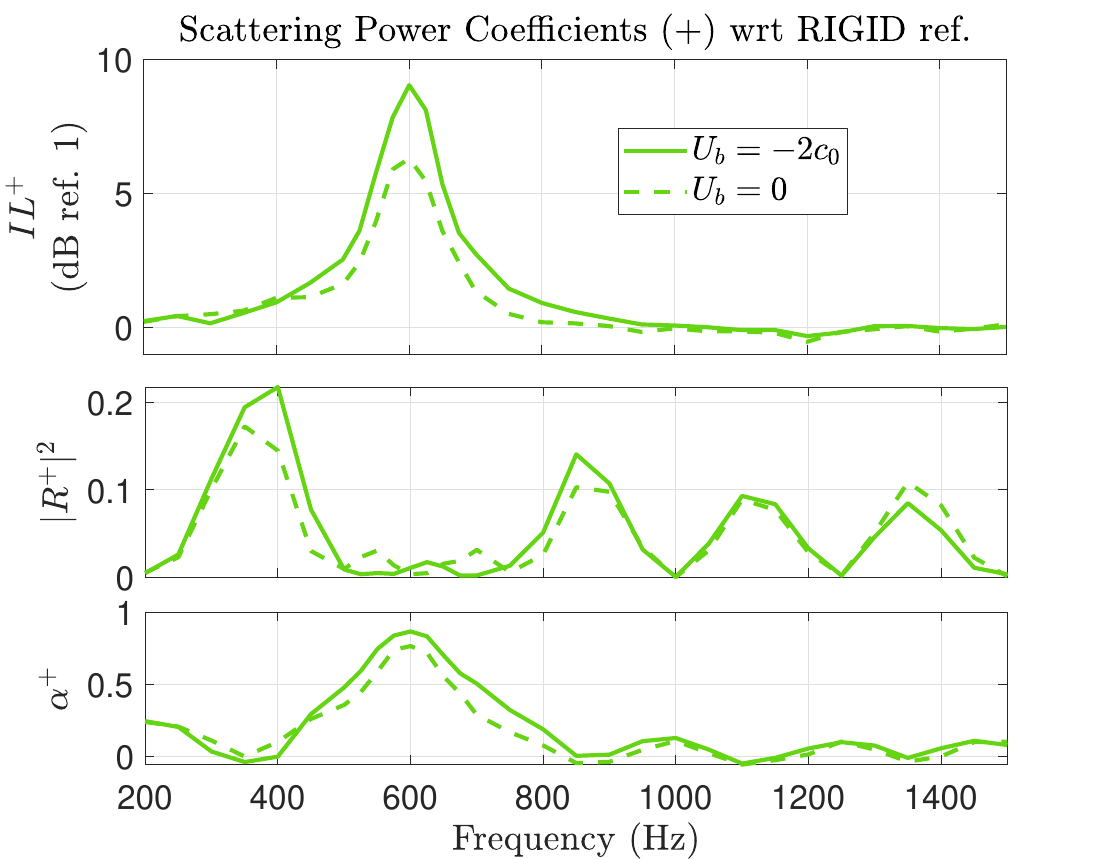}
		\centering
		\caption{}
		\label{fig:ScattPowerCoeffPLUS_05muM_05muK_025rat_VaryingUb_ClassicCells_1500RPM}
	\end{subfigure}
	%	\hfill
	\hspace{1 cm}
	\begin{subfigure}[ht!]{0.45\textwidth}
		\centering
		\includegraphics[width=\textwidth]{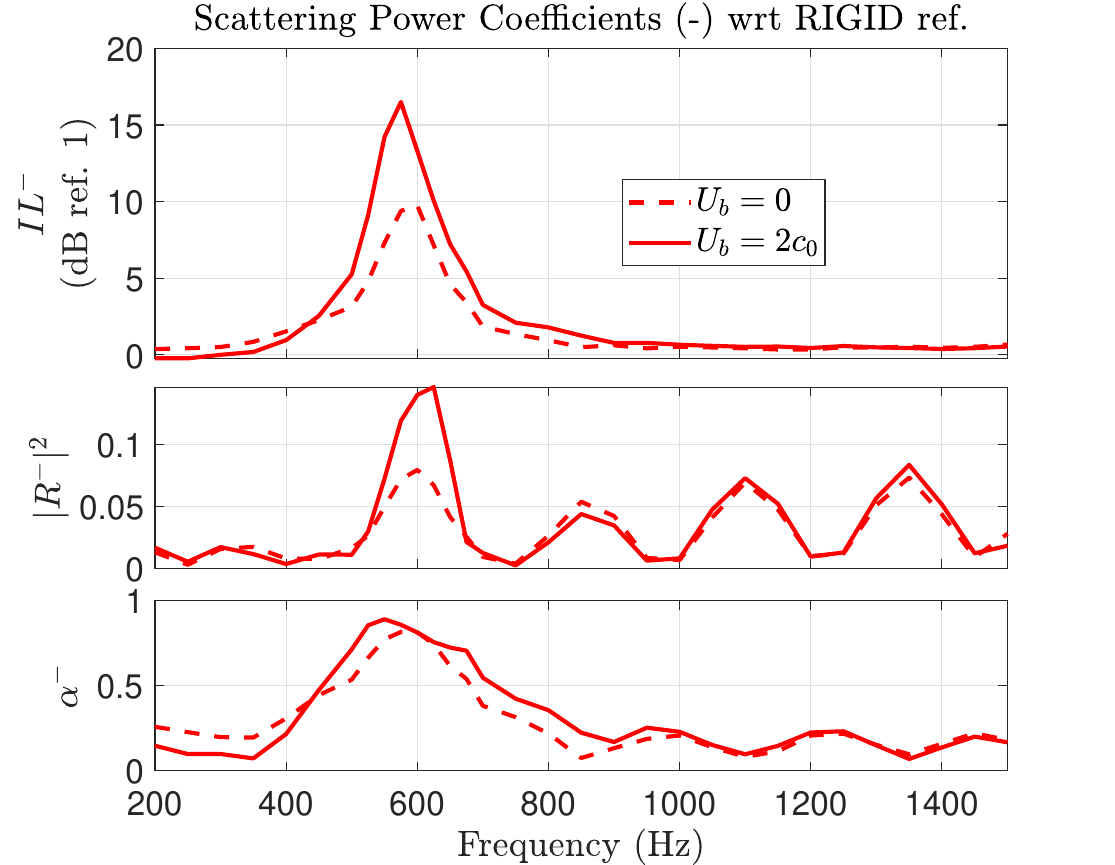}
		\caption{}
		\label{fig:ScattPowerCoeffMINUS_05muM_05muK_025rat_VaryingUb_ClassicCells_1500RPM}
	\end{subfigure}
	\caption{Scattering power coefficients with respect to the rigid reference, in case of Mach 0.15, and targeted resonance frequency 600 Hz, for downstream ($+$) and upstream ($-$) senses of propagation. Both local impedance control ($U_b=0$) and ABL ($|U_b|=2c_0$) performances are reported.}
	\label{fig:Scattering_Caiman_M015}
\end{figure}

%\begin{figure}[ht!]
%	\centering
%	\begin{subfigure}[ht!]{0.45\textwidth}
%		\centering
%		\includegraphics[width=\textwidth]{ScattPowerCoeffPLUS_05muM_05muK_025rat_VaryingUb_3000RPM.pdf}
%		\centering
%		\caption{}
%		\label{fig:ScattPowerCoeffPLUS_05muM_05muK_025rat_VaryingUb_3000RPM}
%	\end{subfigure}
%	%	\hfill
%	\hspace{1 cm}
%	\begin{subfigure}[ht!]{0.45\textwidth}
%		\centering
%		\includegraphics[width=\textwidth]{ScattPowerCoeffPMINUS_05muM_05muK_025rat_VaryingUb_3000RPM.pdf}
%		\caption{}
%		\label{fig:ScattPowerCoeffPMINUS_05muM_05muK_025rat_VaryingUb_3000RPM}
%	\end{subfigure}
%	\caption{$IL$ with respect to the rigid reference, in case of Mach 0.3, and targeted resonance frequency 600 Hz, for downstream ($+$) and upstream ($-$) senses of propagation. Both local impedance control ($U_b=0$) and ABL ($|U_b|=2c_0$) performances are reported.}
%	\label{fig:Scattering_Caiman_M03}
%\end{figure}

In case of mean-flow, we must distinguish the downstream from the upstream propagation. In Fig. \ref{fig:Scattering_Caiman_M015}), we report the scattering performances achieved for the downstream (Fig. \ref{fig:ScattPowerCoeffPLUS_05muM_05muK_025rat_VaryingUb_ClassicCells_1500RPM}) and upstream sense of propagation (Fig. \ref{fig:ScattPowerCoeffMINUS_05muM_05muK_025rat_VaryingUb_ClassicCells_1500RPM}), when Mach is 0.15. The ABL still demonstrates higher isolation performances with respect to local impedance control, for both upstream and downstream propagation. Observe how, in case of mean-flow, the advection speed amplitude on the boundary $|U_b|$ is increased to $2c_0$, with opposite sign with respect to the sense of propagation to attenuate. This is so, in order to efficiently oppose the sound propagation in the waveguide convected by a Mach 0.15. Notice also, that the upstream isolation is enhanced by the air-flow convection in the waveguide, as expected \cite{rice1979modal}.\\

%When the Mach number reaches 0.3 (Fig. \ref{fig:Scattering_Caiman_M03}), the isolation performances in the downstream sense (Fig. \ref{fig:ScattPowerCoeffPLUS_05muM_05muK_025rat_VaryingUb_3000RPM}) no longer show a significant difference between the local and the nonlocal approaches. The reasons are probably to be found in the effect of the wiremesh in front of the electroacoustic liner, which alters the interaction between the programmed boundary and the acoustic domain, and reduce the effectiveness of the ABL. Therefore, the ABL struggles to counteract the natural convection imposed by the air flow in the duct, and reduce the downstream transmission.
%Nevertheless, for the upstream propagation (Fig. \ref{fig:ScattPowerCoeffPMINUS_05muM_05muK_025rat_VaryingUb_3000RPM}), the ABL is still capable of enhancing the isolation performances with respect to the purely local impedance control strategy. This should be kept in mind when confronting with the acoustic treatment at the intake of the turbofan nacelle, where upstream sound radiation is concerned.

\FloatBarrier

\section{Conclusions}

In this paper, we have reported the results which have assessed the isolation performances, robustness and stability of the electroacoustic liner technology, first without flow (both numerically and experimentally), and then with flow. The Advection Boundary Law keeps its higher effectiveness with respect to classical local impedance operators. This considerations open to the second part of this work, related to the assessment of the electroacoustic liner performances to reduce the noise radiation from the intake of a turbofan nacelle reproduction.

\section*{Acknowledgments}
The SALUTE project has received funding from the Clean Sky 2 Joint Undertaking under the European Union’s
Horizon 2020 research and innovation programme under grant agreement N 821093. This publication reflects
only the author’s view and the JU is not responsible for any use that may be made of the information it contains.

\FloatBarrier

\bibliographystyle{plain}
%\bibliographystyle{unsrt}
%\bibliographystyle{model1-num-names}
%\bibliography{biblio_2024}	

%\bibliographystyle{unsrt}

\end{document}